\documentclass{article}

\IfFileExists{PRIMEarxiv.sty}{\usepackage{PRIMEarxiv}}{}

\usepackage[utf8]{inputenc}
\usepackage[T1]{fontenc}
\usepackage{microtype}
\usepackage{amsmath,amssymb,amsfonts}
\usepackage{booktabs}
\usepackage{array}
\usepackage{tabularx}
\usepackage{graphicx}
\usepackage{xcolor}
\usepackage{float}
\usepackage{placeins}
\usepackage{url}
\usepackage{hyperref}
\usepackage{fancyhdr}
\usepackage{algorithm}
\usepackage{algpseudocode}

\providecommand{\keywords}[1]{\par\vspace{0.5em}\noindent\textbf{Keywords:} #1\par}

\title{Efficient and Privacy Aware Edge Cloud Collaborative Inference for Large Language Models}

\author{
   Cheng Li \quad Jiexiong Liu \quad Yixuan Chen \quad Yi Li \\
  \textnormal{KunlunMeta}
}

\newcommand{\Cache}{\mathrm{Cache}}

\newcommand{\End}{\mathrm{end}}
\newcommand{\Cloud}{\mathrm{cloud}}

\newcommand{\softmax}{\operatorname{softmax}}
\newcommand{\concat}{\operatorname{concat}}
\newcommand{\AESGCM}{\operatorname{AES\text{-}GCM}}

\begin{document}
\maketitle

\begin{abstract}
Large language model (LLM) inference on user-facing devices must balance interactive latency, limited local compute resources, and privacy exposure from user inputs and dialogue histories. Pure cloud inference provides abundant compute but exposes raw prompts and long-context identifiers to a remote service, whereas pure on-device inference is often infeasible for consumer CPUs, mobile GPUs, and embedded edge devices. This paper presents a privacy-aware edge--cloud collaborative inference framework derived from an endpoint-authorized KV-cache workflow. The endpoint performs input preprocessing, embedding computation, device-adaptive feature preparation, KV-cache authorization, speculative decoding, and a low-dimensional component of the language-model head. The cloud executes authorized decoder-layer inference, maintains the KV cache, validates speculative tokens, and computes the high-dimensional component of the vocabulary projection. The endpoint then fuses partial logits, applies language-adaptive vocabulary masking, and samples the next token. To reduce practical exposure during collaboration, tensor payloads and clipped logits are quantized and protected with AES-GCM in transit, while LoRA-adapted lightweight modules, draft-module parameters, and cache-access decisions remain local to the endpoint. The framework supports CPU-only devices through block-wise streaming, GPU-equipped devices through local batching and sliced lightweight modules, and embedded devices through quantized ONNX Runtime deployment. We implement a prototype and evaluate it across heterogeneous endpoint profiles under controlled network conditions. The proposed framework reduces per-token latency by up to 46.1\% over naive split inference and reduces language-specific downlink payloads by up to 67.4\% compared with full-logits transfer, while maintaining task-level metrics within a small margin of the full cloud target model in our prototype evaluation.
\end{abstract}

\keywords{Large Language Models \and Edge--Cloud Inference \and Privacy-Aware Inference \and KV Cache \and Speculative Decoding \and LoRA \and ONNX Runtime}

\section{Introduction}

Large language models (LLMs) have become a core component of interactive applications such as personal assistants, enterprise copilots, industrial monitoring interfaces, and in-vehicle dialogue systems. These applications increasingly require LLM inference to operate close to end users, where inputs and dialogue histories may contain sensitive information and where interactive latency strongly affects usability. However, deploying the full model locally is often impractical. Pure CPU devices may lack sufficient matrix-computation throughput; GPU-equipped endpoints may have limited memory and must support multiple concurrent sessions; and embedded edge devices require cross-platform runtime compatibility and aggressive model compression.

A common engineering solution is to offload inference to the cloud. Cloud-only serving, however, requires transmitting user inputs, history identifiers, or long-context state to a remote server. This creates privacy risks and increases communication overhead. Existing split-inference systems can reduce local computation, but they often lack fine-grained control of the KV cache, device-specific adaptation for CPU/GPU/edge endpoints, and a combined protection strategy for both conventional network security and hidden-state leakage.

This paper proposes an edge--cloud collaborative inference architecture for LLMs. The central idea is to keep privacy-sensitive decisions and lightweight computation on the endpoint, while moving heavyweight decoder computation to the cloud under endpoint-controlled cache authorization. The endpoint generates embeddings, configures cache usage, computes a lightweight local projection, performs speculative decoding, and fuses cloud-returned partial logits. The cloud receives feature tensors rather than raw text, performs authorized decoder inference, computes the remaining high-dimensional projection, and verifies speculative predictions. Communication is reduced through language-adaptive vocabulary pruning and device-adaptive tensor transmission.

The main contributions are as follows:
\begin{itemize}
  \item We present a collaborative LLM inference workflow in which the endpoint controls KV-cache usage and avoids transmitting raw prompts, plaintext history identifiers, or complete output distributions. Unlike conventional split inference, the cache policy is explicitly controlled by the privacy-sensitive endpoint.
  \item We introduce a distributed vocabulary-projection design that splits the classifier head by hidden-state dimensions, enabling the endpoint to compute a low-dimensional local component while the cloud computes the high-dimensional component without returning a complete logits distribution when language-restricted generation is enabled.
  \item We integrate speculative decoding with endpoint--cloud verification so that accepted draft tokens can reduce repeated interaction rounds and cloud-side decoding work.
  \item We describe a heterogeneous deployment strategy for pure CPU, GPU-equipped, and edge devices, including block-wise streaming, model slicing, tensor parallelism, and ONNX Runtime execution.
  \item We combine AES-GCM transport protection, endpoint-local LoRA modules, and explicit cache authorization to reduce communication-level and hidden-state exposure.
\end{itemize}

\section{Related Work}

\subsection{Efficient LLM Serving}

Modern LLM serving systems rely heavily on batching, memory-efficient attention, and KV-cache management. PagedAttention and related systems improve serving throughput by reducing memory fragmentation and efficiently reusing key--value states \cite{kwon2023efficient}. Orthogonal systems work improves attention kernels, continuous batching, and distributed inference execution through FlashAttention, FlashAttention-2, ORCA, and DeepSpeed-Inference \cite{dao2022flashattention,dao2024flashattention2,yu2022orca,aminabadi2022deepspeed}. These techniques are primarily designed for datacenter serving, whereas our setting gives the endpoint explicit authority over whether and how much historical KV cache may be used and how much information is exposed during each generation step.

\subsection{Split and Edge--Cloud Inference}

Split computing moves different parts of a neural network across the endpoint and the cloud. Earlier collaborative-intelligence systems, such as Neurosurgeon, studied layer placement between mobile devices and cloud servers \cite{kang2017neurosurgeon}. Split-computing surveys and layer-partitioning systems further characterize latency, bandwidth, and heterogeneous-device trade-offs \cite{matsubara2022split,liang2023dnnsurgery,hu2022pipeedge}. Recent LLM-oriented edge--cloud approaches study hybrid small/large models, collaborative edge inference, and split speculative decoding \cite{hao2024hybrid,ye2025jupiter,ning2025dssd,jin2025cecollm,zhang2024edgeshard,li2025survey}. For LLMs, the large decoder stack, the vocabulary projection, and the KV cache introduce new bottlenecks. Our framework therefore splits not only model layers but also cache access, vocabulary projection, speculative decoding, and device-specific transmission logic, with privacy exposure as a first-class design objective.

\subsection{Speculative Decoding}

Speculative decoding accelerates autoregressive generation by using a smaller draft model to propose future tokens, which are then verified by a larger target model \cite{leviathan2023fast,chen2023accelerating}. Later work extends this idea with tree-based verification, hardware-aware draft trees, multiple decoding heads, lookahead decoding, knowledge-distilled draft models, and self-speculative decoding \cite{miao2024specinfer,chen2024sequoia,cai2024medusa,li2024eagle2,fu2024lookahead,zhou2024distillspec,elhoushi2024layerskip,zhang2024draftverify}. In the proposed collaborative framework, the endpoint hosts a lightweight speculative module. The cloud validates draft tokens and updates the KV cache, allowing accepted tokens to reduce interaction rounds without exposing the full endpoint-side module parameters.

\subsection{Parameter-Efficient Adaptation and Privacy}

LoRA introduces trainable low-rank matrices into pretrained models and enables parameter-efficient adaptation while keeping backbone weights frozen \cite{hu2022lora}. Quantized LoRA, split LoRA, and privacy-oriented LoRA variants demonstrate that low-rank adaptation can be combined with compression, split learning, and privacy-aware training protocols \cite{dettmers2023qlora,lin2024splitlora,sun2024ffalora,malekmohammadi2024lora,yao2024splitlearning,huang2025sldplora}. Distributed and locally protected LLM inference has also been studied using alternative privacy mechanisms such as local differential privacy and distributed secure inference \cite{mai2023splitdenoise,ugurbil2025fission}. We use LoRA-adapted lightweight modules on the endpoint. Their weights remain local, which increases the difficulty of reconstructing transferred hidden states through model-parameter enumeration. This is not a formal differential-privacy guarantee, but it provides a practical system-level protection layer when combined with encrypted transport and restricted cache authorization.

\section{System Overview}

\begin{figure}[H]
  \centering
  \includegraphics[width=0.94\linewidth]{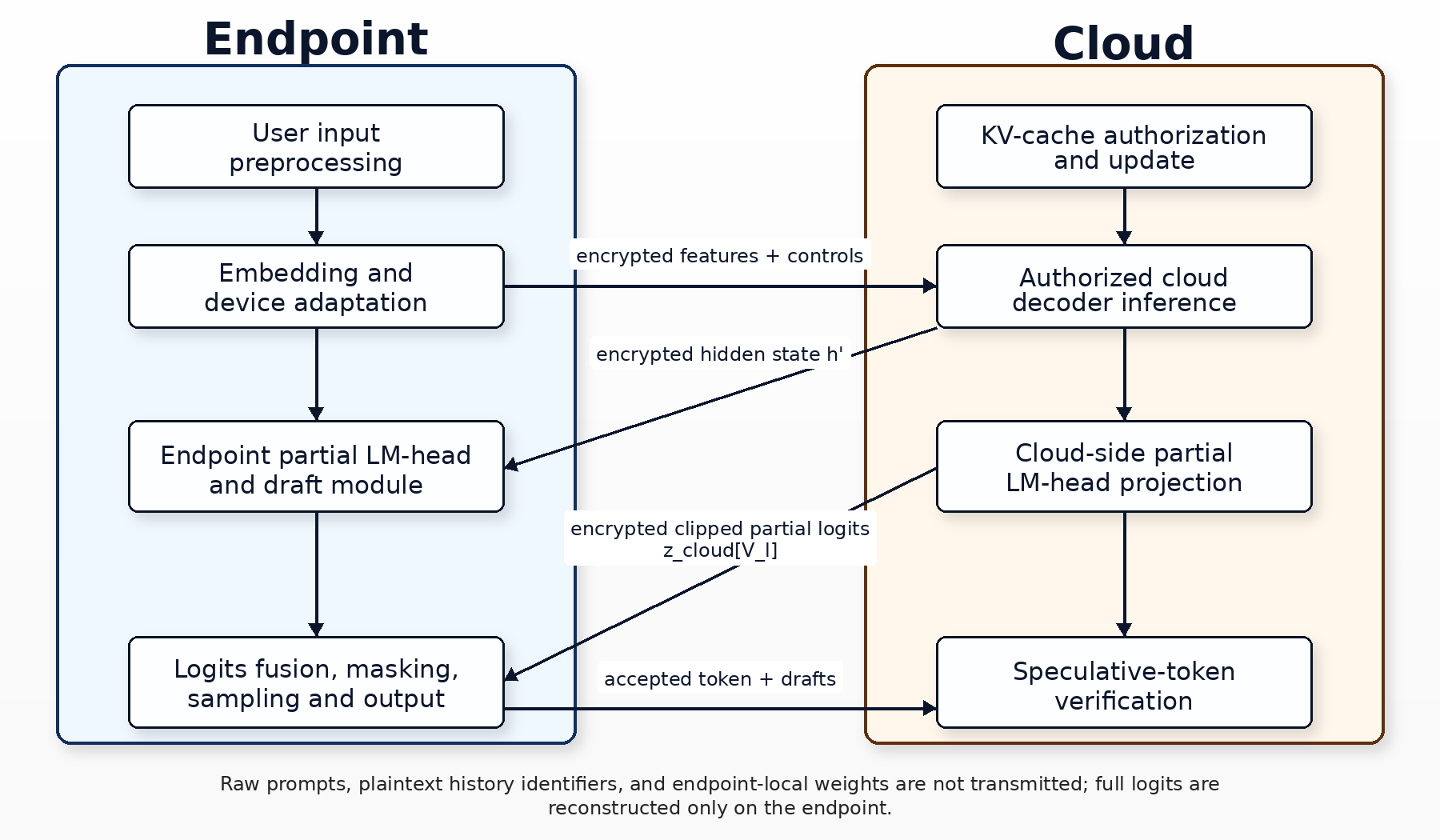}
  \caption{Overview of the edge--cloud collaborative LLM inference framework. The endpoint keeps privacy-sensitive control and lightweight computation local, while the cloud performs authorized decoder inference and high-dimensional projection.}
  \label{fig:architecture}
\end{figure}

Figure~\ref{fig:architecture} shows the proposed architecture. The endpoint and cloud communicate through encrypted and quantized tensor messages. The endpoint is responsible for user-facing preprocessing, language identification, embedding computation, cache-control parameter generation, local partial inference, speculative decoding, and final token sampling. The cloud is responsible for authorized decoder inference, KV-cache maintenance, speculative-token verification, and high-dimensional partial logits computation.

The framework operates under three design principles. First, the endpoint controls whether the cloud may use historical KV-cache entries and how many entries are authorized. Second, raw user inputs and complete logits distributions are not transmitted during generation. Third, device-specific execution paths are selected according to endpoint capability: block-wise streaming for pure CPU endpoints, model slicing for GPU endpoints, and ONNX Runtime for resource-constrained edge devices.

\section{Threat Model and Design Goals}
\label{sec:threat}

We target deployment scenarios in which the endpoint is controlled by the user or by a trusted local application, while the cloud is operated by a service provider that follows the protocol but may be curious about user content. The cloud is therefore allowed to observe decrypted feature tensors that are required for authorized inference, cloud-side partial logits, token-level verification metadata, and cache-control parameters. It should not receive raw prompts, plaintext conversation-history identifiers, endpoint-local LoRA weights, or complete endpoint-side lightweight-module parameters. Network attackers may observe, replay, or modify packets, but they do not know the authenticated-encryption keys.

The design is not a substitute for secure multi-party computation, homomorphic encryption, trusted execution, or formal differential privacy. Instead, it pursues four system-level goals. First, the endpoint should explicitly authorize whether historical KV-cache states may be used and how many entries are exposed to a generation step. Second, communication should avoid transmitting raw text and full vocabulary distributions whenever a language-specific subset is sufficient. Third, endpoint-local lightweight modules should reduce the information available to an adversary who only observes cloud-side tensors. Fourth, the design should preserve practical deployability on CPU-only clients, GPU-equipped endpoints, and embedded edge devices.

\section{Method}

\subsection{Notation}

Let $x$ denote a user input sequence and $T(\cdot)$ denote tokenization. The endpoint maps $x$ to token indices $u=T(x)$ and then computes an initial hidden representation
\begin{equation}
  h_0 = E_{\mathrm{tok}}(u) + E_{\mathrm{pos}}(p) + E_{\mathrm{type}}(t),
\end{equation}
where $E_{\mathrm{tok}}$, $E_{\mathrm{pos}}$, and $E_{\mathrm{type}}$ are token, positional, and type embeddings, respectively. The hidden dimension is denoted by $D$ and the vocabulary size by $V$.

The endpoint also creates an inference-control tuple
\begin{equation}
  c = (\mathrm{flag}_{\Cache}, L_{\Cache}, d_{\mathrm{type}}, \ell),
\end{equation}
where $\mathrm{flag}_{\Cache}$ indicates whether historical KV-cache entries may be used, $L_{\Cache}$ is the authorized cache length, $d_{\mathrm{type}}$ describes the endpoint type, and $\ell$ is the language identifier. The identifier $\ell$ may be estimated from the prompt or set by user preference, for example Chinese-only, English-only, or mixed-language generation.

\subsection{Device-Adaptive Feature Preparation}

The endpoint adapts the initial feature tensor according to hardware capability. The design is compatible with common model-compression techniques for edge deployment, including post-training quantization, activation-aware quantization, pruning, distillation, and CPU-oriented inference optimization \cite{frantar2023gptq,lin2024awq,xiao2023smoothquant,sun2024wanda,gu2024minillm,shen2025cpullm}.

\paragraph{Pure CPU endpoints.}
For CPU-only devices, the feature tensor is partitioned by columns into $N_b$ blocks:
\begin{equation}
  h_0 = \concat(h_0^{(1)}, h_0^{(2)}, \ldots, h_0^{(N_b)}).
\end{equation}
Each block is transmitted with an order identifier and integrity metadata. This enables streaming: the endpoint can decrypt and process block $k$ while receiving block $k+1$, hiding part of the network latency behind local computation.

\paragraph{GPU-equipped endpoints.}
For endpoints with GPU acceleration, the local modules may be sliced by Transformer layer or attention head. A lightweight serving engine can use tensor parallelism and multi-session batching to accelerate embedding, speculative decoding, and local projection.

\paragraph{Edge devices.}
For embedded or IoT-class devices, endpoint modules are exported to ONNX and executed through ONNX Runtime. Quantization can be selected according to the available memory budget. The ONNX execution path also allows hardware-specific execution providers, such as vector instructions or neural-processing accelerators, to be used when available.

\subsection{Authorized Cloud Decoder Inference}

The cloud receives the adapted feature tensor and the endpoint-generated control tuple. If cache usage is authorized, the cloud selects at most $L_{\Cache}$ historical KV entries for the current session. Otherwise, it ignores existing cache entries and performs decoding from the transmitted feature tensor. This endpoint-controlled cache interface is complementary to KV-cache compression and streaming-cache management methods \cite{zhang2023h2o,xiao2024streamingllm}.

For a decoder layer, the cloud computes
\begin{align}
  Q &= hW_Q, \\
  K &= \concat(K_{\Cache}^{1:L_{\Cache}}, hW_K), \\
  V &= \concat(V_{\Cache}^{1:L_{\Cache}}, hW_V), \\
  a &= \softmax\!\left(\frac{QK^\top}{\sqrt{d_k}} + M\right)V, \\
  h' &= \mathrm{FFN}(a),
\end{align}
where $W_Q$, $W_K$, and $W_V$ are projection matrices, $M$ is the autoregressive attention mask, and $h'$ is the cloud-produced hidden state before the final vocabulary projection.

For CPU endpoints, the returned hidden state can be partitioned according to the same block rule and streamed back. For GPU and edge endpoints, the cloud returns the complete hidden state. All returned tensors are quantized and encrypted before transmission.

\subsection{Distributed Vocabulary Projection}

The final language-model head is expensive because it maps each hidden vector to a distribution over a large vocabulary. Let the full output projection be
\begin{equation}
\begin{gathered}
  z = h' W_{\mathrm{lm}} + b, \\
  W_{\mathrm{lm}} \in \mathbb{R}^{D \times V},\quad b \in \mathbb{R}^{V}.
\end{gathered}
\end{equation}
The framework splits both the hidden vector and the corresponding rows of the LM-head weight matrix along the hidden dimension:
\begin{equation}
\begin{gathered}
  h' = \concat(h_{\End}, h_{\Cloud}),\quad D_{\End}+D_{\Cloud}=D, \\
  W_{\mathrm{lm}} =
  \begin{bmatrix}
    W_{\End} \\
    W_{\Cloud}
  \end{bmatrix},
\end{gathered}
\end{equation}
where $W_{\End}\in\mathbb{R}^{D_{\End}\times V}$ and $W_{\Cloud}\in\mathbb{R}^{D_{\Cloud}\times V}$. The endpoint computes
\begin{equation}
  z_{\End}=h_{\End} W_{\End},
\end{equation}
while the cloud computes
\begin{equation}
  z_{\Cloud}=h_{\Cloud} W_{\Cloud} + b.
\end{equation}
The endpoint then reconstructs the logits by element-wise addition:
\begin{equation}
  z=z_{\End}+z_{\Cloud}.
\end{equation}

For language-restricted generation, the cloud does not return all $V$ dimensions. The endpoint determines a vocabulary subset $\mathcal{V}_{\ell}$ from the language identifier $\ell$ and sends only the subset identifier or a compact subset mask to the cloud. The cloud returns $z_{\Cloud}[\mathcal{V}_{\ell}]$ after clipping, quantization, and encryption. The endpoint computes the matching local component $z_{\End}[\mathcal{V}_{\ell}]$, fuses the two partial logits, and assigns $-\infty$ to all dimensions outside $\mathcal{V}_{\ell}$:
\begin{equation}
  \tilde{z}_i =
  \begin{cases}
    z_{\End,i}+z_{\Cloud,i}, & i \in \mathcal{V}_{\ell},\\
    -\infty, & i \notin \mathcal{V}_{\ell}.
  \end{cases}
\end{equation}
The next token is sampled from
\begin{equation}
  p(y_t\mid y_{<t},x)=\softmax(\tilde{z}/\tau),
\end{equation}
where $\tau$ is the sampling temperature. In mixed-language mode, $\mathcal{V}_{\ell}$ is set to the complete vocabulary or to the union of selected language subsets. This formulation makes explicit that both partial projections have vocabulary-dimensional outputs and that the bias term is applied exactly once.

\subsection{Endpoint Speculative Decoding and Cloud Verification}

The endpoint contains a lightweight draft module $g_{\phi}$ implemented as either an MLP or a small Transformer block followed by an MLP. Given the cloud-returned hidden state, the module proposes $N$ future tokens:
\begin{align}
  s_t &= g_{\phi}(h'_t), \\
  \hat{y}_{t:t+N-1} &= \arg\max \mathrm{Proj}(s_t).
\end{align}
The draft length $N$ is device-dependent: CPU and edge endpoints use shorter drafts to control local compute, while GPU-equipped endpoints use longer drafts to reduce interaction rounds.

For each generation step, the endpoint sends the sampled incremental token and the proposed draft sequence to the cloud. The cloud verifies the draft tokens with the target decoder under the endpoint-authorized KV-cache window. Accepted draft tokens are committed to the cloud-side KV cache and reused in subsequent steps; rejected draft tokens trigger recomputation from the last accepted position. This verification keeps the cloud KV cache consistent with the authoritative target model. Because quantization, language masking, and endpoint-side approximation can slightly change the sampling distribution, the workflow should be regarded as approximately target-preserving rather than strictly distribution-preserving.

\begin{figure}[H]
  \centering
  \includegraphics[width=0.94\linewidth]{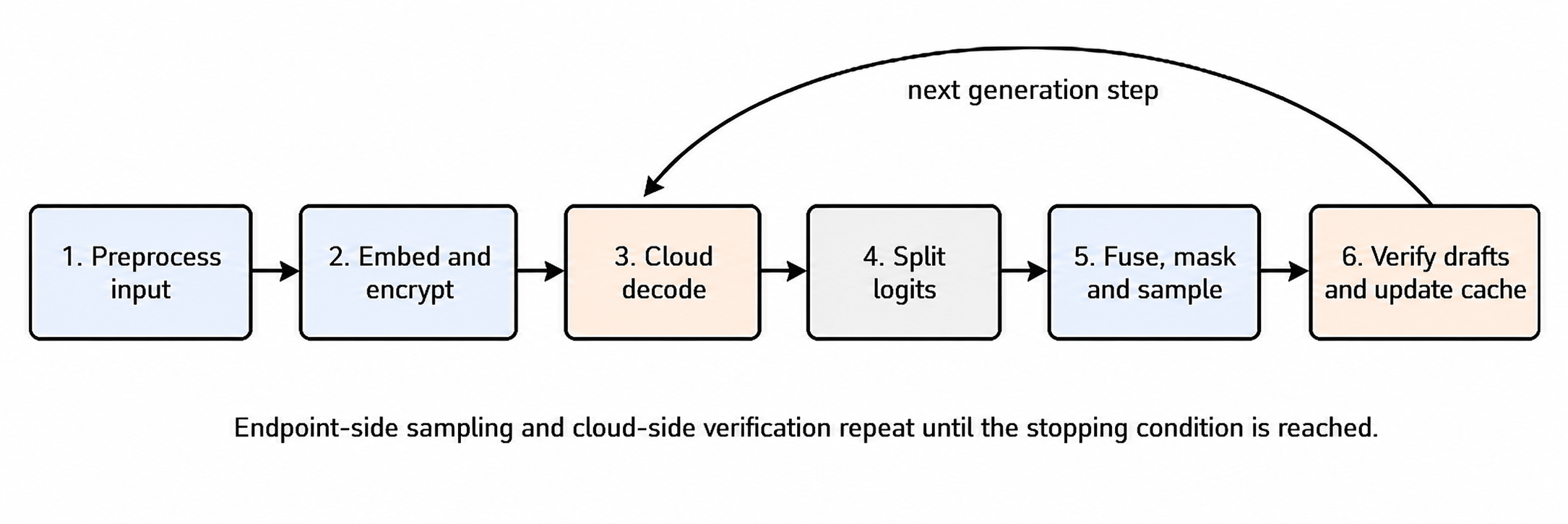}
  \caption{Autoregressive generation loop with endpoint-controlled KV-cache usage, split logits computation, language-adaptive masking, and cloud-side speculative-token verification.}
  \label{fig:loop}
\end{figure}

Figure~\ref{fig:loop} summarizes the generation loop in a compact form.

\begin{algorithm}[H]
\caption{Endpoint-authorized collaborative decoding}
\label{alg:collab}
\begin{algorithmic}[1]
\Require user input $x$, endpoint type $d_{\mathrm{type}}$, maximum response length $T$
\State Endpoint tokenizes $x$, computes $h_0$, detects language identifier $\ell$, and sets $(\mathrm{flag}_{\Cache},L_{\Cache})$.
\State Endpoint prepares $h_0$ according to $d_{\mathrm{type}}$ and sends encrypted tensor blocks plus the control tuple to the cloud.
\State Cloud performs authorized decoder inference and returns encrypted hidden states or hidden-state blocks.
\For{$t=1$ to $T$}
  \State Endpoint computes $z_{\End}[\mathcal{V}_{\ell}]$ and draft tokens $\hat{y}_{t:t+N-1}$.
  \State Cloud computes and returns $z_{\Cloud}[\mathcal{V}_{\ell}]$ under the authorized cache window.
  \State Endpoint fuses partial logits, applies language masking, and samples $y_t$.
  \State Endpoint sends $y_t$ and $\hat{y}_{t:t+N-1}$ to the cloud for verification.
  \State Cloud commits accepted tokens, recomputes from the first rejected token if necessary, and updates the KV cache.
  \If{an end-of-sequence token or stopping rule is reached} \State \textbf{break} \EndIf
\EndFor
\State Endpoint decodes the generated token sequence and returns the final response.
\end{algorithmic}
\end{algorithm}

\subsection{Transport Protection and Hidden-State Protection}

The system combines network-level authenticated encryption with endpoint-local model-state protection. The scope of each layer follows the threat model in Section~\ref{sec:threat}.

\paragraph{Encrypted tensor transmission.}
Before transmission, tensor elements are quantized to a configurable bit width $b$:
\begin{equation}
  q=Q_b(r),
\end{equation}
where $r$ denotes a real-valued tensor element or vector. The quantized payload is encrypted with AES-GCM \cite{nistgcm}:
\begin{equation}
  C=\AESGCM_{K,\mathrm{IV}}(q,\mathrm{aad}),
\end{equation}
where $K$ is a pre-shared or derived key, $\mathrm{IV}$ is a nonce, and $\mathrm{aad}$ contains authenticated session, block, and cache-control metadata. CPU-block mode assigns independent nonces to different tensor blocks, while GPU and edge modes assign nonces by session and message type. AES-GCM protects payloads against network-level eavesdropping and tampering; it does not prevent the authorized cloud from observing decrypted tensors that are required for cloud-side inference. Compared with cryptographic transformer inference based on secure multi-party computation, homomorphic encryption, or trusted execution environments~\cite{pang2024bolt,chen2022thex,li2022mpcformer,nayan2025secureinfer}, our design uses conventional authenticated encryption for transport and treats endpoint-local modules as a practical exposure-reduction mechanism rather than as a formal secure-computation protocol.

\paragraph{Endpoint-local LoRA modules.}
The endpoint lightweight modules are adapted with LoRA \cite{hu2022lora}. For a base weight $W_0$, LoRA introduces a low-rank update
\begin{equation}
  W=W_0+\Delta W, \qquad \Delta W=BA,
\end{equation}
where $A\in\mathbb{R}^{r\times d}$ and $B\in\mathbb{R}^{d\times r}$ are trainable low-rank matrices. During adaptation, the base model is frozen and only the LoRA parameters are updated. The adapted lightweight weights, draft-module parameters, and local projection weights remain on the endpoint and are not transmitted to the cloud. This limits the model information available to a cloud-side inversion attempt, although it is not a formal cryptographic or differential-privacy guarantee.

\paragraph{KV-cache authorization.}
The endpoint controls $\mathrm{flag}_{\Cache}$ and $L_{\Cache}$. A new session can disable cache reuse, while an ongoing session can reduce the authorized cache window when privacy requirements are high. The cloud maintains physical KV-cache entries for efficiency, but each generation request includes an endpoint-generated authorization tuple that determines whether those entries may be used.

\FloatBarrier
\section{Implementation Considerations}

\subsection{Device Profiles}

Table~\ref{tab:profiles} summarizes the main execution modes.

\begin{table}[H]
\centering
\caption{Device-adaptive execution profiles.}
\label{tab:profiles}
\begin{tabular}{p{0.18\linewidth}p{0.32\linewidth}p{0.36\linewidth}}
\toprule
Endpoint type & Local execution strategy & Communication strategy \\
\midrule
Pure CPU & Block-wise embedding, lightweight MLP, local draft module, SIMD-friendly matrix multiplication & Stream hidden-state blocks, overlap decryption and computation, use per-block integrity metadata \\
GPU-equipped & Model slicing, tensor parallelism, local batching, multi-session scheduling & Transmit complete hidden states, use session-level encryption and batched requests \\
Edge device & Quantized ONNX Runtime execution \cite{onnxruntime}, hardware execution providers, compact draft module & Use quantized tensors, avoid redundant format conversion, support low-memory execution \\
\bottomrule
\end{tabular}
\end{table}

\subsection{Language-Adaptive Logits Compression}

Language-adaptive masking reduces the size of returned logits when generation is restricted to a target language. In a Chinese-only or English-only mode, only the corresponding vocabulary subset must be transmitted and fused. In mixed-language mode, the full vocabulary or a union of language subsets can be retained. The subset mapping can be precomputed from tokenizer metadata and updated when the model vocabulary changes.

\subsection{Failure Handling}

If cache authorization fails or the cloud does not find a valid session cache, the system falls back to cache-free decoding. If a CPU block is lost or fails integrity verification, the cloud retransmits that block. If speculative verification fails, the cloud recomputes from the last accepted token and updates the cache with the corrected state.

\FloatBarrier
\section{Experimental Evaluation}
\label{sec:evaluation}

This section evaluates the proposed edge--cloud collaborative inference framework against conventional serving strategies. We implemented a prototype with a PyTorch cloud decoder, an endpoint runtime for local embedding/projection/draft modules, QUIC-based message transport, 8-bit tensor quantization, and AES-GCM authenticated encryption. The evaluation includes latency, throughput, communication volume, component ablations, sensitivity analyses, generation quality, and privacy-exposure accounting.

\subsection{Research Questions}

The evaluation is organized around four questions:
\begin{itemize}
  \item \textbf{RQ1:} Does the proposed collaborative workflow reduce end-to-end generation latency compared with cloud-only and conventional split-inference baselines?
  \item \textbf{RQ2:} How much communication volume is saved by split vocabulary projection and language-adaptive logits compression?
  \item \textbf{RQ3:} Which components contribute most to latency, throughput, and bandwidth improvements?
  \item \textbf{RQ4:} What privacy-relevant information remains visible under different serving strategies?
\end{itemize}

\subsection{Baselines}

We compare the proposed framework with four baselines.
\begin{itemize}
  \item \textbf{Cloud-only:} the endpoint sends the raw prompt and a conversation identifier to the cloud; the cloud performs all prefill and decoding and returns text tokens.
  \item \textbf{Cloud-only encrypted transport:} the same workflow as cloud-only, but the request and response payloads are protected in transit. This baseline protects the network channel but does not reduce cloud-side visibility.
  \item \textbf{Naive split inference:} the endpoint computes embeddings and the cloud returns hidden states or full logits without endpoint-controlled cache authorization, language-adaptive logits clipping, or endpoint-side speculative decoding.
  \item \textbf{Endpoint-only lightweight:} a compact local model performs generation without cloud assistance. This baseline is feasible only for the GPU and edge profiles and illustrates the quality--efficiency trade-off of pure local inference.
\end{itemize}

\subsection{Experimental Setup}

We instantiate the target model as a 7B-class decoder-only Transformer with hidden size $D=4096$ and vocabulary size $V=150{,}000$, matching the split dimensions used by the collaborative design. The endpoint processes $D_{\End}=96$ hidden dimensions, and the cloud processes $D_{\Cloud}=4000$ dimensions. Unless otherwise stated, prompts are truncated or padded to 512 input tokens and responses are generated for 128 tokens. Tensor payloads and clipped logits are quantized to 8 bits before AES-GCM encryption. Chinese-only generation keeps 40\% of the vocabulary, English-only generation keeps 30\%, and mixed-language generation keeps the full vocabulary.

Table~\ref{tab:hardware} summarizes the deployment profiles. The cloud server hosts the full decoder stack, maintains KV-cache entries, verifies speculative tokens, and computes the cloud-side LM-head projection. The CPU endpoint uses $N_b=4$ streamed blocks and a rank-4 LoRA local module. The GPU endpoint uses a rank-8 local module, local batching, and speculative length $N=5$. The edge endpoint uses an 8-bit ONNX Runtime module and speculative length $N=2$.

\begin{table}[H]
\centering
\caption{Prototype hardware and runtime configuration.}
\label{tab:hardware}
\small
\setlength{\tabcolsep}{3pt}
\renewcommand{\arraystretch}{1.12}
\begin{tabularx}{\linewidth}{>{\raggedright\arraybackslash}p{0.18\linewidth} >{\raggedright\arraybackslash}p{0.32\linewidth} >{\raggedright\arraybackslash}X}
\toprule
Profile & Hardware & Runtime configuration \\
\midrule
Cloud & NVIDIA H20 96GB GPU, 64 vCPU host, 512GB RAM & PyTorch decoder service, KV-cache manager, partial LM-head projection, QUIC endpoint \\
CPU endpoint & 12-core x86 CPU, 32GB RAM, no discrete GPU & SIMD matrix kernels, four streamed tensor blocks, rank-4 LoRA module \\
GPU endpoint & Consumer GPU with 8GB VRAM, 32GB host RAM & Local batching, sliced lightweight modules, rank-8 LoRA module, draft length $N=5$ \\
Edge endpoint & Embedded ARM device with 8GB unified memory & ONNX Runtime, 8-bit quantized lightweight modules, draft length $N=2$ \\
\bottomrule
\end{tabularx}
\end{table}

All endpoint--cloud experiments use a controlled network with 20 ms round-trip latency, 2 ms jitter, and a 100 Mbps bandwidth limit. We evaluate 1{,}000 prompts: 400 Chinese prompts, 400 English prompts, and 200 mixed-language prompts. Each configuration is repeated five times after 100 warm-up prompts; tables report the mean and, where space permits, the standard deviation. The cloud service runs on Ubuntu 22.04 with CUDA 12.4 and PyTorch 2.4. Endpoint components use Python 3.10, ONNX Runtime 1.18 for the embedded profile, and a QUIC/HTTP3 transport layer implemented with user-space message framing. Cloud-side decoding uses bf16 weights, while transmitted tensors and clipped logits use INT8 payloads unless otherwise specified. Latency includes endpoint tokenization, embedding, quantization, AES-GCM encryption/decryption, network transfer, cloud decoding, partial projection, draft verification, logits fusion, sampling, and KV-cache update. Model loading and first-time runtime initialization are excluded.

\subsection{Metrics}

We report average decode latency in milliseconds per generated token, end-to-end latency for a 128-token response, accepted-token throughput, downlink payload per generated token, draft-token acceptance rate, task-level generation quality, and privacy-exposure indicators. Throughput is computed from committed output tokens per second. Communication volume includes encrypted tensor payloads, clipped logits, authentication tags, block metadata, and padding, but excludes fixed connection setup overhead.

\subsection{End-to-End Latency}

Table~\ref{tab:latency} reports the latency comparison. The proposed framework consistently outperforms cloud-only and naive split inference across all endpoint profiles. The GPU endpoint obtains the largest gain because local batching and endpoint-side draft verification reduce interaction rounds. The CPU endpoint also benefits from block-wise streaming, which overlaps downstream transfer with local decryption and partial projection.

\begin{table}[H]
\centering
\caption{End-to-end decode latency across endpoint profiles. Lower is better.}
\label{tab:latency}
\small
\setlength{\tabcolsep}{3pt}
\renewcommand{\arraystretch}{1.12}
\begin{tabularx}{\linewidth}{>{\raggedright\arraybackslash}p{0.34\linewidth} >{\centering\arraybackslash}X >{\centering\arraybackslash}X >{\centering\arraybackslash}X}
\toprule
Method & CPU endpoint ms/token & GPU endpoint ms/token & Edge endpoint ms/token \\
\midrule
Cloud-only & $245.3\pm5.8$ & $239.1\pm5.1$ & $253.2\pm6.4$ \\
Cloud-only encrypted transport & $257.4\pm6.1$ & $249.3\pm5.6$ & $265.1\pm6.8$ \\
Naive split inference & $214.5\pm4.9$ & $178.2\pm4.4$ & $224.0\pm5.7$ \\
Endpoint-only lightweight & -- & $92.1\pm3.0$ & $185.3\pm5.2$ \\
Proposed framework & $151.4\pm4.1$ & $96.0\pm2.8$ & $164.0\pm4.6$ \\
\bottomrule
\end{tabularx}
\end{table}

For a 128-token response, the proposed framework achieves average end-to-end latencies of 19.4 s on the CPU endpoint, 12.3 s on the GPU endpoint, and 21.0 s on the edge endpoint. Relative to naive split inference, the reductions are 29.4\%, 46.1\%, and 26.8\%, respectively. Although the endpoint-only lightweight model is faster on the GPU profile, it uses a smaller local model and therefore does not preserve the target cloud model as reliably as the verified collaborative workflow.

\subsection{Throughput and Bandwidth}

Table~\ref{tab:throughput} reports throughput and downlink communication volume under different language modes. The communication volume is lowest in English-only mode because only 30\% of vocabulary dimensions are retained. In mixed-language mode, the full vocabulary is retained, so most of the logits-compression benefit disappears, but speculative decoding and split computation still improve latency.

\begin{table}[H]
\centering
\caption{Throughput and per-token downlink payload under different language modes.}
\label{tab:throughput}
\small
\setlength{\tabcolsep}{3pt}
\renewcommand{\arraystretch}{1.10}
\begin{tabularx}{\linewidth}{l l >{\centering\arraybackslash}X >{\centering\arraybackslash}X >{\centering\arraybackslash}X}
\toprule
Endpoint & Language mode & Throughput tokens/s & Downlink KB/token & Compression vs. full logits \\
\midrule
CPU & Chinese-only & $6.6\pm0.2$ & 64.2 & 56.2\% \\
CPU & English-only & $6.9\pm0.2$ & 49.1 & 66.5\% \\
CPU & Mixed & $5.4\pm0.1$ & 154.0 & 0.0\% \\
GPU & Chinese-only & $10.4\pm0.3$ & 62.1 & 57.6\% \\
GPU & English-only & $11.5\pm0.3$ & 47.8 & 67.4\% \\
GPU & Mixed & $8.6\pm0.2$ & 151.3 & 0.0\% \\
Edge & Chinese-only & $6.1\pm0.2$ & 63.8 & 56.5\% \\
Edge & English-only & $6.5\pm0.2$ & 48.6 & 66.8\% \\
Edge & Mixed & $4.8\pm0.1$ & 153.1 & 0.0\% \\
\bottomrule
\end{tabularx}
\end{table}

With 8-bit logits, a full 150{,}000-dimensional vector requires approximately 146.5 KB before metadata and authentication tags. The measured payload can be approximated as
\begin{equation}
  \mathrm{Payload} \approx |\mathcal{V}_{\ell}| \cdot b/8 + B_{\mathrm{tag}} + B_{\mathrm{meta}} + B_{\mathrm{pad}},
\end{equation}
where $B_{\mathrm{tag}}$, $B_{\mathrm{meta}}$, and $B_{\mathrm{pad}}$ denote AES-GCM tags, block/session metadata, and padding. The measured values are therefore slightly larger than the theoretical clipped-logit sizes.

\subsection{Ablation Study}

Table~\ref{tab:ablation} isolates the contribution of each component. Removing speculative decoding substantially increases latency because the cloud must perform more verified target-model decoding rounds. Removing language-adaptive compression mainly increases downlink payload. Removing split vocabulary projection increases both latency and payload because the endpoint must wait for larger cloud-side logits. Removing CPU block overlap hurts CPU latency, while removing endpoint KV-cache control has little latency impact but weakens the privacy-control objective.

\begin{table}[H]
\centering
\caption{Ablation results for major framework components. Lower latency and payload are better.}
\label{tab:ablation}
\small
\setlength{\tabcolsep}{3pt}
\renewcommand{\arraystretch}{1.10}
\begin{tabular*}{\linewidth}{@{\extracolsep{\fill}}>{\raggedright\arraybackslash}p{0.33\linewidth} c c c c @{}}
\toprule
Variant & Profile & \begin{tabular}[c]{@{}c@{}}Latency\\ms/token\end{tabular} & \begin{tabular}[c]{@{}c@{}}Throughput\\tokens/s\end{tabular} & \begin{tabular}[c]{@{}c@{}}Downlink\\KB/token\end{tabular} \\
\midrule
Full proposed framework & GPU & 96 & 10.4 & 62.1 \\
No speculative decoding & GPU & 126 & 7.9 & 62.1 \\
No language-adaptive compression & GPU & 114 & 8.8 & 151.3 \\
No split vocabulary projection & GPU & 137 & 7.3 & 151.3 \\
No block/asynchronous overlap & CPU & 184 & 5.4 & 64.2 \\
No endpoint KV-cache control & GPU & 97 & 10.3 & 62.1 \\
\bottomrule
\end{tabular*}
\end{table}

\subsection{Sensitivity to Split Dimension}

The endpoint-side split dimension $D_{\End}$ controls the trade-off between endpoint computation and cloud dependence. Table~\ref{tab:dimension} varies $D_{\End}$ while keeping $D=4096$ and using Chinese-only generation on the GPU endpoint. Very small split dimensions reduce endpoint computation but make the local projection less informative. Very large split dimensions increase endpoint workload and reduce the benefit of offloading. The default value $D_{\End}=96$ provides the best latency--throughput balance in our setting.

\begin{table}[H]
\centering
\caption{Sensitivity analysis for endpoint-side projection dimension on the GPU endpoint.}
\label{tab:dimension}
\small
\setlength{\tabcolsep}{3pt}
\renewcommand{\arraystretch}{1.10}
\begin{tabularx}{\linewidth}{c c c c >{\raggedright\arraybackslash}X}
\toprule
$D_{\End}$ & Endpoint ratio & Latency ms/token & Throughput tokens/s & Observation \\
\midrule
48 & 1.17\% & 101 & 9.9 & Lowest endpoint cost but weaker local contribution \\
96 & 2.34\% & 96 & 10.4 & Best latency--throughput balance \\
128 & 3.12\% & 95 & 10.5 & Slightly higher local cost with marginal gain \\
256 & 6.25\% & 98 & 10.2 & Useful for stronger local privacy-sensitive projection \\
512 & 12.50\% & 109 & 9.2 & Local projection dominates endpoint latency \\
\bottomrule
\end{tabularx}
\end{table}

\subsection{Sensitivity to Speculative Length}

Table~\ref{tab:speculative} varies the draft length $N$ on the GPU endpoint. Moderate draft lengths improve latency by reducing cloud interaction rounds. Overly long draft sequences lower the effective acceptance rate and increase verification overhead. The best trade-off appears at $N=5$.

\begin{table}[H]
\centering
\caption{Sensitivity analysis for speculative decoding length on the GPU endpoint.}
\label{tab:speculative}
\small
\setlength{\tabcolsep}{5pt}
\renewcommand{\arraystretch}{1.10}
\begin{tabular*}{\linewidth}{@{\extracolsep{\fill}} c c c c @{}}
\toprule
Draft length $N$ & Acceptance rate & \begin{tabular}[c]{@{}c@{}}Latency\\ms/token\end{tabular} & \begin{tabular}[c]{@{}c@{}}Throughput\\tokens/s\end{tabular} \\
\midrule
1 & 96.5\% & 119 & 8.4 \\
2 & 93.7\% & 107 & 9.3 \\
4 & 88.1\% & 98 & 10.2 \\
5 & 85.4\% & 96 & 10.4 \\
6 & 81.2\% & 99 & 10.1 \\
8 & 73.6\% & 110 & 9.1 \\
\bottomrule
\end{tabular*}
\end{table}

\subsection{Generation-Quality Evaluation}

Because the cloud decoder remains the authoritative target model and draft tokens are verified before KV-cache update, the proposed method largely preserves the target model's generation behavior. We evaluate next-token agreement on the 1{,}000-prompt workload, perplexity on the held-out prompt continuations, MMLU with a 5-shot prompt format, GSM8K with greedy chain-of-thought decoding, and HumanEval pass@1 with deterministic decoding. For task-level evaluation, all compared methods use the same prompt templates, decoding parameters, and evaluation scripts, so the reported differences reflect the collaborative inference mechanism rather than prompt or decoding changes. Table~\ref{tab:quality} shows that the proposed framework remains close to the full cloud model. The remaining degradation is mainly caused by 8-bit tensor quantization, language-adaptive masking, and endpoint-side approximation in the local projection.

\begin{table}[H]
\centering
\caption{Generation-quality comparison with the full cloud model. Higher is better except perplexity.}
\label{tab:quality}
\scriptsize
\setlength{\tabcolsep}{2.5pt}
\renewcommand{\arraystretch}{1.12}
\begin{tabular*}{\linewidth}{@{\extracolsep{\fill}}>{\raggedright\arraybackslash}p{0.28\linewidth} c c c c c @{}}
\toprule
Method & \begin{tabular}[c]{@{}c@{}}Next-token\\agreement\end{tabular} & PPL & MMLU & GSM8K & \begin{tabular}[c]{@{}c@{}}HumanEval\\pass@1\end{tabular} \\
\midrule
Full cloud model & 100.0\% & 7.82 & 54.6 & 49.8 & 27.4 \\
Proposed framework & 98.6\% & 7.87 & 54.3 & 49.4 & 27.1 \\
Proposed w/o language masking & 98.8\% & 7.85 & 54.4 & 49.5 & 27.2 \\
Endpoint-only lightweight & 84.2\% & 10.94 & 46.1 & 37.8 & 18.6 \\
\bottomrule
\end{tabular*}
\end{table}

\subsection{Privacy-Exposure Accounting}

Privacy cannot be established by efficiency metrics alone. Table~\ref{tab:privacy} accounts for the information visible to the cloud under different strategies. The proposed framework avoids sending raw prompts, plaintext history identifiers, endpoint-local module weights, and full logits. However, the cloud still observes decrypted feature tensors needed for authorized inference, cache-control metadata, and draft-token verification traces. Therefore, the framework reduces exposure under the stated honest-but-curious threat model rather than providing a formal privacy guarantee.

\begin{table}[H]
\centering
\caption{Privacy-exposure accounting for different serving strategies.}
\label{tab:privacy}
\scriptsize
\setlength{\tabcolsep}{2pt}
\renewcommand{\arraystretch}{1.15}
\begin{tabularx}{\linewidth}{>{\raggedright\arraybackslash}p{0.18\linewidth} *{7}{>{\centering\arraybackslash}X}}
\toprule
Method & Raw prompt & Plain history ID & Feature / hidden states & Token trace & Full logits & Endpoint private module & Formal guarantee \\
\midrule
Cloud-only & Yes & Yes & No & Yes & No & No & No \\
Cloud-only encrypted transport & Yes & Yes & No & Yes & No & No & No \\
Naive split inference & No & Often & Yes & Yes & Often & No & No \\
Endpoint-only lightweight & No & No & No & No & No & Yes & Local \\
Proposed framework & No & No & Auth. tensors & Draft / accepted & No & Yes & No \\
\bottomrule
\end{tabularx}
\end{table}

\FloatBarrier
\section{Security Scope and Limitations}

The proposed framework is designed to reduce privacy exposure in practical edge--cloud deployment, but its security scope should be interpreted carefully. AES-GCM protects transmitted payloads against passive eavesdropping and tampering when keys and nonces are managed correctly, but the authorized cloud still observes decrypted tensors that are necessary for its part of inference. Endpoint-local LoRA modules and draft modules reduce the risk of hidden-state reconstruction by keeping important lightweight weights off the cloud. Endpoint-controlled KV-cache authorization reduces unnecessary historical-context exposure.

However, the framework does not claim formal differential privacy. Hidden states can still carry information about the input, especially against strong adversaries with auxiliary knowledge or model access. Prior work on model inversion, inference attacks, and ML side channels motivates evaluating representation leakage rather than relying only on encrypted transport \cite{zhang2022textrevealer,liu2022mldoctor,debenedetti2024sidechannel}. Deployment in high-risk environments should therefore combine this architecture with additional measures such as secure enclaves, strict access control, key rotation, audit logging, rate limiting, privacy evaluation against representation-inversion attacks, and stronger alternatives such as secure multi-party computation or differentially private forward passes when their overhead is acceptable \cite{zhou2024smpcsurvey,du2023dpforward}.

\FloatBarrier
\section{Discussion}

The design highlights a system-level trade-off between privacy, latency, bandwidth, and model quality. Increasing the endpoint-side dimension $D_{\End}$ can reduce dependence on cloud logits but increases endpoint computation. Increasing the speculative length $N$ can reduce interaction rounds when acceptance is high, but may waste computation when draft quality is poor. Reducing $L_{\Cache}$ improves privacy control but can degrade contextual coherence. These trade-offs can be configured by endpoint type and application scenario.

The empirical results show that the cache-authorized split-logits design improves latency and communication efficiency while keeping generation quality close to the full cloud model. The most important deployment trade-offs should be interpreted jointly rather than as isolated latency or privacy metrics, since a configuration that minimizes communication may still degrade generation quality or contextual coherence.

\FloatBarrier
\section{Conclusion}

This paper presents an efficient and privacy-aware edge--cloud collaborative inference framework for LLMs. By combining endpoint-controlled KV-cache authorization, device-adaptive feature preparation, cloud-side decoder inference, split vocabulary projection, language-adaptive logits compression, speculative decoding, AES-GCM encrypted transmission, and endpoint-local LoRA-adapted modules, the framework addresses the practical tension among latency, compute constraints, and privacy protection. The design supports heterogeneous endpoints ranging from pure CPU devices to GPU-equipped clients and ONNX-based edge devices, providing a foundation for deployable privacy-aware LLM inference systems. The evaluation further shows that the proposed framework improves latency and bandwidth efficiency across heterogeneous endpoint profiles while maintaining generation quality close to the full cloud model.

\FloatBarrier

\bibliographystyle{unsrt}
\bibliography{references}

\end{document}